\documentclass[aps,groupedaddress,showpacs,amsmath,amssymb,twocolumn]{revtex4-1}

\usepackage{graphicx}
\usepackage{bm}
\usepackage{wasysym}
\usepackage{natbib}
\usepackage[usenames]{xcolor}

\begin{document}

\title{Tunable axial potentials for atom-chip waveguides}

\author{James A. Stickney}
\email{james.stickney@sdl.usu.edu}
\affiliation{Space Dynamics Laboratory, North Logan, Utah 84341, USA}

\author{Brian Kasch}
\affiliation{Air Force Research Laboratory, Kirtland AFB, New Mexico 87117, USA}

\author{Eric Imhof}
\affiliation{Space Dynamics Laboratory, North Logan, Utah 84341, USA}

\author{Bethany Kroese}
\author{Jonathon A.R. Crow}
\author{Spencer E. Olson}
\author{Matthew B. Squires}
\affiliation{Air Force Research Laboratory, Kirtland AFB, New Mexico 87117, USA}

\date{\today}

\begin{abstract}

We present a method for generating precise, dynamically tunable magnetic potentials that can be described by a polynomial series along the axis of a cold-atom waveguide near the surface of an atom chip.  With a single chip design consisting of several wire pairs, various axial potentials can be created by changing the ratio of the currents in the wires, including double wells, triple wells, and pure harmonic traps with suppression of higher-order terms.  We use this method to design and fabricate a chip with modest experimental requirements.  Finally, we use the chip to demonstrate a double-well potential.  

\end{abstract}


\maketitle

\section{INTRODUCTION}{\label{sec:Intro}}

Experiments with cold atoms often rely on carefully designed magnetic fields to create potentials for specific experimental requirements. A leading method for generating magnetic potentials involves the design and fabrication of ``atom chips''~\cite{keil_fifteen_2016, fortagh_magnetic_2007}, conducting wires on one or more dielectric substrates. 
Typically, the trap shape is determined by the wire pattern while the magnitude of the trapping field is determined by the chip currents. This enables a broad range of possible magnetic trap positions and parameters. Magnetic trap capabilities can be expanded with the addition of radio frequency~\cite{schumm_matter-wave_2005,Hofferberth2007-1} and microwave~\cite{Bohi2009} fields. Periodic wire structures~\cite{grabowski_lattice_2003, yun_practical_2006, yin_magnetic_2002}, permanent magnets~\cite{hall_condensate_2007}, diffractive magnetic lattices~\cite{gunther_combined_2005, gunther_diffraction_2005}, and optical elements for the generation~\cite{imhof_two-dimensional_2017, mcgilligan_grating_2017, cotter_design_2016}, manipulation~\cite{Du2004}, and detection~\cite{colombe_strong_2007} of ultracold ensembles have been successfully integrated with atom chips.
Due to their extensive configurability, and compact size, atom chips have become a cornerstone of emerging atomic sensor technologies~\cite{Carter2013,berrada_integrated_2013,Chuang2011,Robins2013,abend_atom-chip_2016,tino_precision_2013}.

\begin{figure}
\includegraphics[width=.8 \columnwidth]{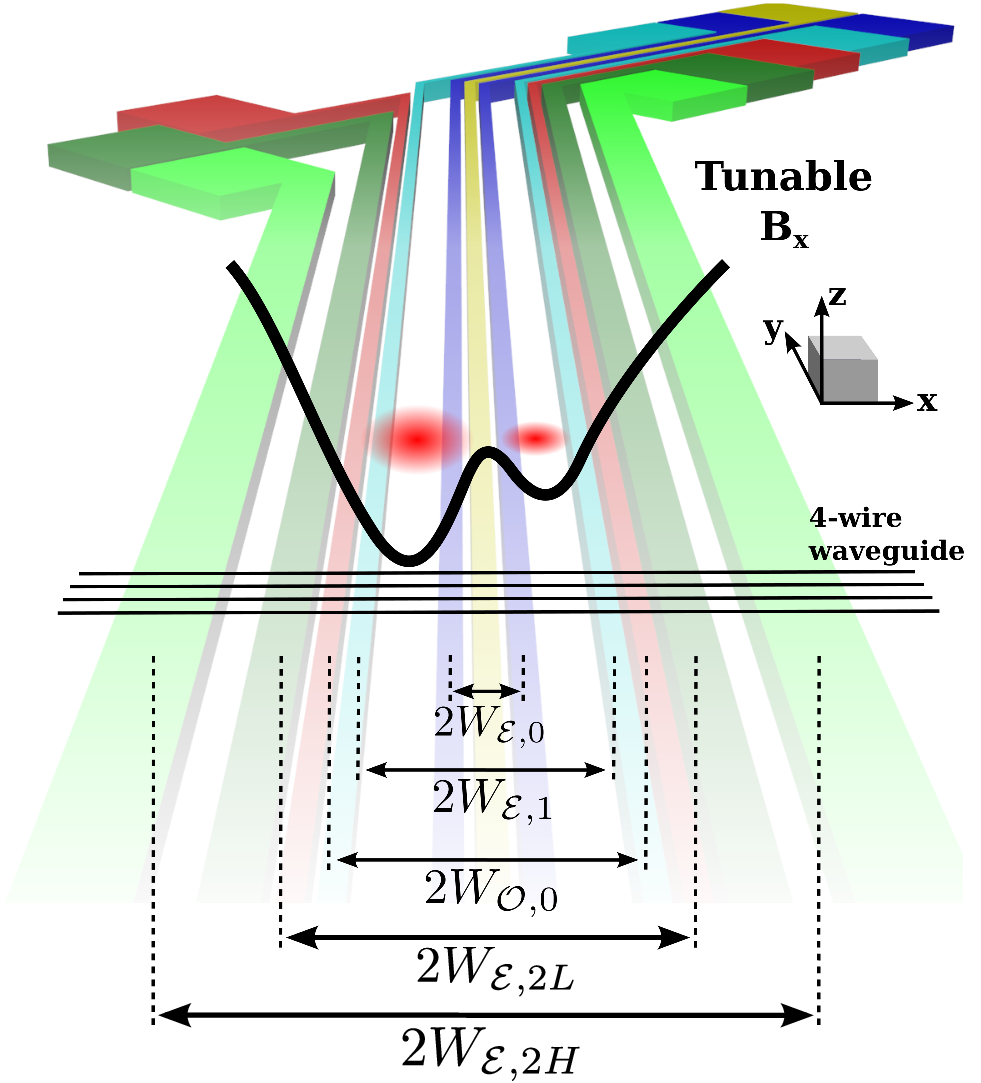}
\caption{Multilayer atom chip with tunable control over $B_{x}$ along a cold-atom waveguide. The black crossing wires on the top layer are used to form a four-wire waveguide. Below, even and odd wire pairs, spaced by $2W_{\mathcal{P},m}$, give control over even and odd contributions to the 1D potential.
\label{fig:TuneableTrapWires}}
\end{figure}

In this paper, we present a design methodology for producing dynamically tunable one-dimensional (1D) magnetic potentials by summing the magnetic-field contributions from multiple wire pairs on an atom chip. 
The field of a single wire can be Taylor expanded about the position of the atom trap, revealing a polynomial series. With two wires of equal or opposite current, equidistant from the location of the trap minimum, odd or even terms, respectively, are eliminated from the series expansion due to symmetry. With multiple wire pairs of the appropriate spacing, an orthogonal basis set is realized. The linear combination of the fields creates a total magnetic field that can be approximated as an $n$th-order polynomial in the trapping region. 
We show that the order of the polynomial is determined by the number and locations of the wire pairs and that the coefficients of the polynomial are determined by the ratio of currents in the wire pairs.  

Although fabrication methods for atom chips vary considerably, from standard milling of metallic films~\cite{vale_foil-based_2004}, to optical and $e$-beam lithography~\cite{Trinker2008} to electro-chemical~\cite{Squires2011} and laser etching~\cite{squires_ex_2016} of direct bonded copper on an aluminum nitride substrate, our technique is architecture independent. The chip design is parameterized only by the working distance between the atoms and the chip and the desired number of tunable orders in the polynomial expansion of the field. 

Since both even and odd contributions are accessible in a single chip design, highly arbitrary polynomial potentials can be realized utilizing a single layer of chip conductors.  
This tunability applies to a variety of experiments, including trapped-atom interferometry~\cite{Burke2009, Westbrook2009, wang_atom_2005}, chip-based precision measurements~\cite{Riedel2010,abend_atom-chip_2016}, 1D Bose gases~\cite{Jacqmin2012}, and atomtronic devices~\cite{Chuang2011,Zozulya2013,Hansel2001}.

In practice, canceling higher-order terms requires a higher power dissipation. Therefore, we consider a low-power wire configuration that relaxes the requirements on higher-order terms. Based on the calculations outlined in this paper, we have designed, fabricated, and tested an atom chip capable of controlling the 1D potential including both the optimal and reduced-power wire configurations.

The rest of the paper is organized as follows. In Sec.~\ref{sec:OneDPotential} we present an idealized atom chip and its corresponding 1D polynomial potential and examine the tunability of both even and odd terms. In Sec.~\ref{sec:Example}, we describe an example chip design and solve for wire currents using either the optimal or the low-power configuration. In Sec.~\ref{sec:Conclusions} we present initial experimental results showing the tunability of the potential and summarize our findings.

\section{Magnetic-field control in one dimension \label{sec:OneDPotential} }

In many experiments, an atom cloud is sufficiently confined in two directions such that its dynamics can be described by a 1D equation of motion.
In this paper, a radial plus an effective 1D axial potential is formed by pairs of wires patterned on a two-layer atom chip
with an adjustable uniform external magnetic field. 
This chip is shown schematically in Fig.~\ref{fig:TuneableTrapWires}.
The layer closest to the atoms will be used to create a magnetic waveguide~\cite{Thywissen1999}, depicted as a set of four horizontal black wires, that tightly confines the atoms in the radial directions.
The far layer is composed of multiple wire pairs which create a tunable axial field perpendicular to the waveguide. The wire pairs will be referred to as pinch wires since they act much like the pinch coils in a Ioffe trap. While Fig.~\ref{fig:TuneableTrapWires} shows finite wires with leads, the following derivation assumes infinitely long thin wires. 

In Appendix~\ref{sec:oneDField}, we show the axial and radial potentials are separable when  $\mu |B_x(0)| \gg m \omega_{\perp}^2 \sigma_{\perp}^2$, where $\sigma_{\perp}$ is the characteristic size of the atomic cloud in the radial direction and $B_x(0)$ represents the total bottom field in the waveguide.  The effective 1D axial potential along $x$ can then be written as
\begin{equation}
  \label{eq:Ve}
  V = \mu | B_x(x) | + \frac{1}{2} m \omega_{\perp}^2 r_{\perp}^{2},
\end{equation}
where $\mu=\mu_B g_F m_F$ is the magnetic moment of the atomic state that is trapped, $\mu_B$ is the Bohr magneton, $g_F$ is the Land\'{e} $g$-factor, $m_F$ is the magnetic quantum number,  $B_x(x)$ is the tunable magnetic field in the $\hat e_x$ direction, $\omega_{\perp}$ is the trapping frequency in the radial direction, $r_{\perp}$ is the distance from the trap axis, and $m$ is the atomic mass.  

\begin{figure}[!th]
\includegraphics[width=\columnwidth]{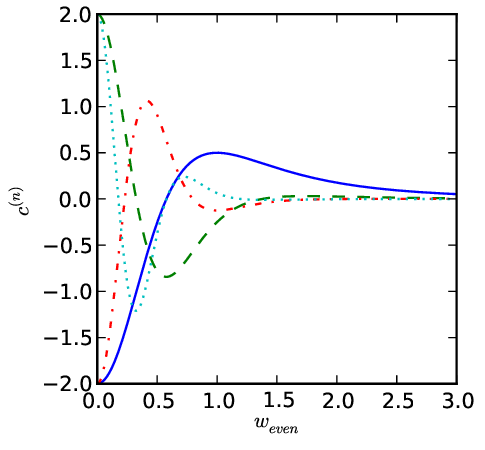}
\caption{\label{fig:cn_even} 
The lowest few even coefficients, $c^{(n)}$ are shown.  The solid line is $c^{(2)}$, the dashed line is $c^{(4)}$, the dash-dotted line is $c^{(6)}$, and the dotted line is $c^{(8)}$. }
\end{figure}

For a single wire pair centered about the origin with both currents running in the $\hat e_y$ direction the field can be expressed as the following series:
\begin{equation}
  \begin{split}
  \label{eq:little_c_x}
  B_x(x) =& \frac{\mu_0 I}{2 \pi H} \left[ c^{(0)}  +  c^{(2)} \left( \frac{x}{H} \right)^2 + c^{(4)} \left( \frac{x}{H} \right)^4 + \ldots \right],
  \end{split}
\end{equation}
where $I$ is the current in the wire pair and $H$ is the distance of the atom trap from the plane of the 1D control wires. In Appendix~\ref{sec:GenExp}, we show that the parameters $c^{(n)}$ are given by the relation
\begin{equation}
  \label{eq:cn}
  c^{(n)}(w) = \frac{2}{(1+w^2)^{n+1}} \sum_{r=0}^{n} (-1)^{(n+r)/2} \binom{n+1}{r} w^{r} \phi_{n+r},
\end{equation}
for any $n$, where $\phi_{a} = [1+(-1)^a]/2$ is a parity function of the integer argument $a$ that is $1$ if $a$ is even and $0$ if $a$ is odd.  Figure ~\ref{fig:cn_even} shows the first few even values of the coefficients $c^{(n)}$, given in Eq.~(\ref{eq:cn}), as a function of half the scaled wire spacing $w_{\mathcal{E}} = W_{\mathcal{E}}/H$, where $W_{\mathcal{E}}$ is half the real wire spacing between a wire pair on the chip.  The odd orders cancel due to the symmetry of the wire spacing and the currents.  For antisymmetric current flow there is a Taylor series similar to Eq.~(\ref{eq:little_c_x}) except the even orders cancel such that there are only odd terms.  Equation (\ref{eq:cn}) holds for both even and odd contributions to the potential.

Akin to a Helmholtz or anti-Helmholtz coil pair there is a particular wire spacing where one of the orders in the expansion will cancel.   When just one wire pair is flowing current, scaling this current will equally scale all orders of the series expansion, but does not change the functional form of the potential. However, if the atom chip is designed with the appropriate wire-pair spacings, individual terms in the total tunable field can be varied by changing the relative currents in the wire pairs.   
We use multiple wire pairs at various spacings such that the potential is a linear combination of the Taylor series expansions of each wire pair. This can be contrasted with previous studies which examine the magnetic field of various coil configurations~\cite{turner_gradient_1993, crow_method_1996, garrett_axially_1951, boridy_magnetic_1989, hyodo_mirror_2007, jollenbeck_hexapole-compensated_2011, bergeman_magnetostatic_1987}, or analogously, the electric field of an arrangement of charged electrodes~\cite{gabrielse_open-endcap_1989}, by performing a multipole expansion in the region of interest.

\begin{figure}[!th]
\includegraphics[width=\columnwidth]{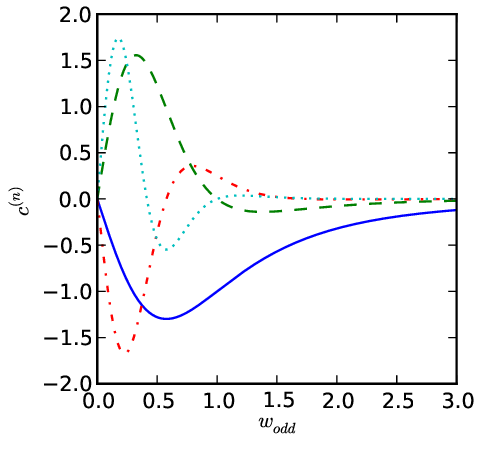}
\caption{
\label{fig:cn_odd}
The lowest few odd coefficients, $c^{(n)}$ are shown.  The solid line is $c^{(1)}$, the dashed line is $c^{(3)}$, the dash-dotted line is $c^{(5)}$,
and the dotted line is $c^{(7)}$.
}
\end{figure}

The total tunable field can be expanded into the following series (note that the $c$ terms denote the magnetic-field contribution from a single wire pair, while $C$ denotes a sum of $c$ terms):
\begin{equation}
  \begin{split}
  \label{eq:beta_x}
  B_x(x) =& B^{*}_{x} + B_R \left[ C^{(0)} + C^{(1)} \left( \frac{x}{H} \right) +  C^{(2)} \left( \frac{x}{H} \right)^2 + \right. \\ & \left. C^{(3)} \left( \frac{x}{H} \right)^3 + C^{(4)} \left( \frac{x}{H} \right)^4 + \ldots \right],
  \end{split}
\end{equation}
where
\begin{equation}
  \label{eq:Cn_even}
  C^{(n)} = \sum^{M(\mathcal{P})-1}_{m=0} i_{\mathcal{P},m} c^{(n)}(w_{\mathcal{P},m}).
\end{equation}
where $i_{\mathcal{P},m} = I_{\mathcal{P},m}/I_R$ is the relative current in the $m$th wire pair and $w_{\mathcal{P},m} = W_{\mathcal{P},m}/H$, scaled by the distance from the waveguide axis to the wire plane. The parity of $n$ determines which wire pairs contribute to $C^{(n)}$.  The number of contributing wire pairs is given by $M(\mathcal{P})$, where $\mathcal{P}$ denotes the parity of $n$, denoted either $\mathcal{E}$ for even, or $\mathcal{O}$ for odd.
Additionally, $B^{*}_{x}$ is the magnitude of the externally applied, uniform bias field in the $x$ direction, and $B_R$ is the overall potential scaling given by
\begin{equation}
  \label{eq:beta_0}
  B_R = \frac{\mu_0 I_R}{2 \pi H},
\end{equation}
where $I_R$ is the reference current.  

The rest of this section describes how arbitrary values of the $C^{(n)}$ coefficients can be generated from a particular wire configuration.

To ensure separability we compute the bottom field by summing the externally applied field with terms from the even wire pairs,
\begin{equation}
  \label{eq:B0}
  B_x(0) = B^{*}_{x} + B_R \sum^{M(\mathcal{E})-1}_{m=0} i_{\mathcal{E},m} c^{(0)}(w_{\mathcal{E},m}).
\end{equation}
Note that the odd wires make no contribution to the bottom field. 

The magnetic field of the wire pairs also consists of a component in the $z$ direction.
This field can be expanded as
\begin{equation}
  \label{eq:Bz}
  B_z(x) = B^{*}_{z} + B_R \left(D^{(0)} + D^{(1)} \frac{x}{H} + \ldots \right).
\end{equation}
The opposite parity condition in the $z$ direction means that the dimensionless parameters $D^{(n)}$ are determined by the currents in wires of parity $n+1$, of which there are $M(\mathcal{P}^{\prime})$. These coefficients are given by,
\begin{equation}
  \label{eq:Dn}
  D^{(n)} = \sum^{M(\mathcal{P}^{\prime})-1}_{m=0}i_{\mathcal{P}^{\prime},m} d^{(n)}(w_{\mathcal{P}^{\prime},m}),
\end{equation}
where $d^{(n)}$ are dimensionless parameters that depend only on the spacing of the wires.  

The parameter $D^{(0)}$ causes a displacement of the wave guide in the $z$ direction.  However, this constant field can be corrected with the addition of a uniform bias field $B^{*}_{z}$.  In the rest of the paper we assume that the correct bias field is applied.  

Non zero values of $D^{(1)}$ cause a rotation of the waveguide.  Typically, this rotation is set to zero; however, there are situations where changing this rotation angle will be useful, such as the alignment of a cloud with a standing-wave laser field.  Extensions to non zero rotations are straightforward but will be neglected in what follows.  

A general expression for $d^{(n)}$, similar to the one given in Appendix~\ref{sec:GenExp}, may be found.  However, we are interested in only the two lowest orders, which can be expressed as
\begin{equation}
   \label{eq:d0}
   d^{(0)}(w_{\mathcal{O},m}) = 2 \frac{w_{\mathcal{O},m}}{w_{\mathcal{O},m}^2+1},
 \end{equation}
and
\begin{equation}
  \label{eq:d1}
  d^{(1)}(w_{\mathcal{E},m}) = -2 \frac{(w_{\mathcal{E},m}-1)(w_{\mathcal{E},m}+1)}{(w_{\mathcal{E},m}^2+1)^2}.
\end{equation}

The currents in a set of $M(\mathcal{E})$ wire pairs can be used to control usually the lowest $M(\mathcal{E})-1$ terms from Eq.~(\ref{eq:Cn_even}) plus the parameter $D^{(1)}$ from Eq.~(\ref{eq:Dn}).  For a given set of wire spacings $\{w_{\mathcal{E}}\}$ the currents can be found by inverting Eqs.~(\ref{eq:Cn_even}) and~(\ref{eq:Dn}).  Once the currents have been found, the contributions to the potential from the uncontrolled parameters can be calculated. 

By placing wire pairs at the roots of a coefficient, we can eliminate the contribution to the potential from that coefficient. As Fig.~\ref{fig:cn_even} shows, each even coefficient has one more zero crossing than the previous one; that is, $c^{(2)}$ has one root, $c^{(4)}$ has two roots, $c^{(6)}$ has three roots, etc.  Thus, an atom chip can be designed to produce a polynomial of any even order with the next highest order being exactly canceled.

The number of roots is exactly the number of wires needed to control all of the lower coefficients plus $D^{(1)}$. By placing wire pairs at all of the roots of a given even coefficient and controlling the relative current through each pair, one can tune the lower even coefficients, as well as the additional coefficient $D^{(1)}$. For example, by placing wires at the three roots of $c^{(6)}$, we can independently control the three parameters $C^{(2)}$, $C^{(4)}$, and $D^{(1)}$, while also having $C^{(6)} = 0$.    

Similarly, the currents in a set of  $M(\mathcal{O})$ wire pairs can be used to control $M(\mathcal{O})$ terms from Eq.~(\ref{eq:Cn_even}).  Once the wire spacings $\{w_{\mathcal{O}}\}$ have been determined, the currents are found by inverting Eqs.~(\ref{eq:Cn_even}) and~(\ref{eq:Dn}).  
An applied bias field of $B^{*}_{z} = -B_RD^{(0)}$ is required to cancel the $D^{(0)}$ that arises from the odd wire pairs. The value can be calculated from the currents, and using Eqs.~(\ref{eq:Dn}) and~(\ref{eq:d0}).

Figure ~\ref{fig:cn_odd} shows the first few odd values of the coefficients $c^{(n)}$, given in Eq.~(\ref{eq:cn}), as a function of half the wire spacing $w_{\mathcal{O}} = W_{\mathcal{O}}/H$.  The solid line is $c^{(1)}$, the dashed line is $c^{(3)}$, the dash-dotted line is $c^{(5)}$, and the dotted line is $c^{(7)}$.
Like with the even case, each of the odd coefficients has one more root than the previous one.  However, one of the roots is always at $w_{\mathcal{O}} = 0$.  This root cannot be used to create an odd potential and is therefore not useful.  As a result, $c^{(1)}$ has no useful roots, $c^{(3)}$ has one useful root, $c^{(5)}$ has two useful roots, etc.  

By placing the odd wires at the useful roots of a coefficient, all of the lower coefficients can be controlled.  For example, by placing wires at the two roots of $c^{{(5)}}$, the coefficients $C^{(1)}$ and $C^{(3)}$ can be controlled, and $C^{(5)} = 0$.  The dominant contribution of the $z$ component of the field $D^{{(0)}}$ can be eliminated using a bias field.  It is not necessary to have a wire pair to control its value.

\section{Examples}{\label{sec:Example}}

We will first determine the placement of the wire pairs and then describe two potentials that can be generated with the design.  Consider the case of three even wire pairs and two odd wire pairs.  These wires can be used to create any potential that is described by a fourth-order polynomial.  Once the coefficients and wire spacings are specified, the set of currents $\{i_{\mathcal{P}}\}$ can be found by solving the following matrix equations. For the even wires,
\begin{equation}
  \label{eq:matrix_equationEven}
  \begin{split}
  & \left(
\begin{array}{ccc}
c^{(2)}(w_{\mathcal{E},0}) & c^{(2)}(w_{\mathcal{E},1}) & c^{(2)}(w_{\mathcal{E},2}) \\
c^{(4)}(w_{\mathcal{E},0}) & c^{(4)}(w_{\mathcal{E},1}) & c^{(4)}(w_{\mathcal{E},2}) \\
d^{(1)}(w_{\mathcal{E},0}) & d^{(1)}(w_{\mathcal{E},1}) & d^{(1)}(w_{\mathcal{E},2}) \\
\end{array}
 \right) \left(
\begin{array}{c}
i_{\mathcal{E},0} \\
i_{\mathcal{E},1} \\
i_{\mathcal{E},2}
\end{array}
\right) \\ &= \left(
\begin{array}{c}
C^{(2)} \\
C^{(4)} \\
0
\end{array}
\right),
\end{split}
\end{equation}
and for odd wires,
\begin{equation}
  \label{eq:matrix_equationOdd}
  \left(
\begin{array}{ccc}
c^{(1)}(w_{\mathcal{O},0}) & c^{(1)}(w_{\mathcal{O},1}) \\
c^{(3)}(w_{\mathcal{O},0}) & c^{(3)}(w_{\mathcal{O},1}) \\
\end{array}
 \right) \left(
\begin{array}{c}
i_{\mathcal{O},0} \\
i_{\mathcal{O},1}
\end{array}
\right) = \left(
\begin{array}{c}
C^{(1)} \\
C^{(3)} 
\end{array}
\right).
\end{equation}
Equations.~(\ref{eq:matrix_equationEven}) and~(\ref{eq:matrix_equationOdd}) can be used to set the coefficients $C^{(1)}$ through $C^{(4)}$ for any given wire spacing.  However, contributions to the higher-order terms of the potential will generally depend on these wire spacings.  

The sixth-order contribution can be eliminated, $C^{(6)} = 0$, by placing the wires with spacing of $w_{\mathcal{E},0} = 0.228$, $w_{\mathcal{E},1} = 0.797$, and $w_{\mathcal{E},2H} = 2.076$.  The fifth-order contribution is always zero, $C^{(5)} = 0$, when $w_{\mathcal{O},0} = 0.577$ and $w_{\mathcal{O},1} = 1.732$.

In situations where small sixth-order contributions to the potential can be tolerated, the total power consumption of the atom chip can be greatly reduced by moving the outer pair of wires closer together.  We choose to place the outer wires at a spacing where $w_{\mathcal{E},2} = w_{\mathcal{E},2L} = 1.3$.  This choice has much lower power requirements than the optimal spacing while maintaining a rather low contribution from the sixth-order term.

Several example trap configurations will now be discussed.  In all cases, we utilize an atom chip with a working distance of $H = 1.6$~mm between the atoms and the central plane of the tuning wires. There are four free parameters, $C^{(1)}$ through $C^{(4)}$.  For each trap type, the results will be presented for the optimal configuration where $w_{\mathcal{E},2} = w_{\mathcal{E},2H} = 2.076$ and a low-power configuration where $w_{\mathcal{E},2} = w_{\mathcal{E},2L} = 1.3$.  For both of these configurations, it will be assumed that the odd wires are placed to provide optimal tuning of the odd coefficients, as discussed above. 

For the optimal and low-power configurations, Eqs.~(\ref{eq:matrix_equationEven}) and~(\ref{eq:matrix_equationOdd}) are numerically inverted.  For the case of the optimal configuration the currents are given by the relations
\begin{equation}
\label{highPowerInverted}
\left(
\begin{array}{c}
i_{\mathcal{E},0} \\
i_{\mathcal{E},1} \\
i_{\mathcal{E},2H} 
\end{array}
\right)
=
\left(
\begin{array}{ccc}
 1.33 & 1.47 & 1.05 \\
 2.31 & 0.70 & 1.64 \\
 12.37 & 11.54 & 5.31
\end{array}
\right) 
\left(
\begin{array}{c}
C^{(2)} \\
C^{(4)} \\
D^{(1)} 
\end{array}
\right),
\end{equation}
and for the low-power configuration
\begin{equation}
\label{eq:lowPowerInverted}
\left(
\begin{array}{c}
i_{\mathcal{E},0} \\
i_{\mathcal{E},1} \\
i_{\mathcal{E},2L} 
\end{array}
\right)
= 
\left(
\begin{array}{ccc}
 0.31 & 0.52 & 0.61\\
 0.29 & -1.18 & 0.77\\
 3.18 & 2.96 & 1.36
\end{array}
\right)
\left(
\begin{array}{c}
C^{(2)} \\
C^{(4)} \\
D^{(1)} 
\end{array}
\right).
\end{equation}
With one exception, the magnitude of the currents in the high-power configuration (\ref{highPowerInverted}) is always larger than the values in the low power configuration.  This is especially true of the last row in the matrices, which determines the current in the outer most wire.  

Assuming the resistances of each of the pinch wires are equal, the total power dissipated is given as the sum of the squares of the currents.  For the harmonic potential in the optimal configuration, the power dissipation is proportional to $\sum_m i_m^2 = 160.12$, and in the low-power configuration the power dissipation is proportional to $\sum_m i_m^2 = 10.29$.  For the case of a harmonic potential, the power dissipation due to the pinch wires is 15 times less for the low-power configuration.  In addition, the low-power configuration requires a smaller external bias.
We are interested in the case where $D^{(1)} = 0$, so the last row in both matrices will not be used in the discussion that follows below.  

The inverted equation for the odd terms is 
\begin{equation}
\left(
\begin{array}{c}
i_{\mathcal{O},0} \\
i_{\mathcal{O},1} 
\end{array}
\right)
= 
\left(
\begin{array}{cc}
 -0.19 & 0.77 \\ 
 -1.73 & -2.31 
\end{array}
\right)
\left(
\begin{array}{c}
C^{(1)} \\
C^{(3)} 
\end{array}
\right).
\end{equation}
Finally, the bias field needed to cancel the $D^{{(0)}}$ term is 
\begin{equation}
  \label{eq:Bz_star}
  B^{*}_z / B_R = 1.66 C^{(1)} + 1.33 C^{(3)}.
\end{equation} 

\subsection{Harmonic trap}

This tunable trap will be useful for atom interferometry in harmonic traps.  This is particularly true for an interferometer that uses trapped thermal atoms, because contributions to the fourth- (and higher-) order term cause decoherence due to the larger size of the cloud which samples more of the potential.  Additionally, higher-order contributions to the potential can be caused by the finite length of chip wires, the leads that connect the chip wires to the power supplies, ion pumps, or other laboratory equipment.  
These contributions can be canceled by tuning the parameter $C^{(4)}$, while holding $C^{(2)}$ constant. $C^{(4)}$ can be tuned both positive and negative to cancel any stray $C^{(4)}$ coefficient.   To effectively remove the effects of the fourth-order contributions to the potential, the background value of $C^{(4)}$ must first be determined.  We are currently developing methods for measuring these fourth-order contributions and plan on using the chip described in this paper to evaluate the effectiveness of these methods.  

\begin{figure}
\includegraphics[width=\columnwidth]{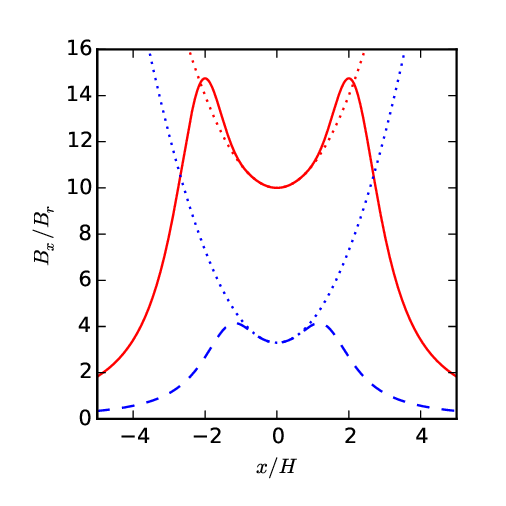}
\caption{{With $C^{(2)} = 1$ and all higher-order terms zeroed, the field along $x$ is harmonic. The solid red curve shows the field produced by the wires in the optimal configuration $w_{\mathcal{E},2} = w_{\mathcal{E},2H}$. The dashed blue curve shows the field produced by the wires in the low-power configuration $w_{\mathcal{E},2} = w_{\mathcal{E},2L}$. Dotted lines represent idealized harmonic field profiles. } \label{Fig:harmonic}}
\end{figure}
Before tuning the parameter $C^{(4)}$, a harmonic trap must first be created and loaded.  Figure ~\ref{Fig:harmonic} shows the magnetic field for the case where $C^{(2)} = 1$ and all other coefficients are zero.  The solid red curve shows the field produced by the wires in the optimal configuration $w_{\mathcal{E},2} = w_{\mathcal{E},2H}$, and the dashed blue curve shows the field produced by the wires in the low-power configuration $w_{\mathcal{E},2} = w_{\mathcal{E},2L}$.  The dotted lines show the field profile when higher-order terms are neglected.  With the pinch wires in the optimal configuration, the trap remains harmonic over a larger range.  The optimal trap is also deeper and has a larger bottom field.  Thus, the bias field to reduce the bottom field will need to be larger for the optimal configuration as compared to the low-power configuration.  

\begin{figure}
\includegraphics[width=\columnwidth]{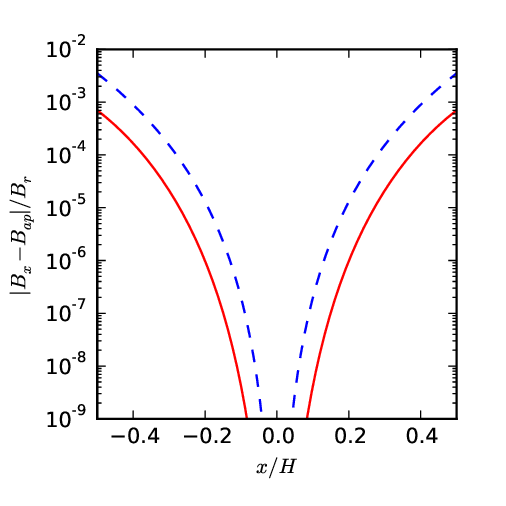}
\caption{{Comparison of the deviations of the axial field from pure harmonicity for optimal configuration versus the low-power configuration. The high-power configuration (solid red line) results in an order-of-magnitude improved harmonicity over the low-power configuration (dashed blue line).} \label{Fig:harmonicLog}}
\end{figure}

To quantify the effects of the uncontrolled higher-order contributions of the field, Fig.~\ref{Fig:harmonicLog} shows a log plot of the difference between the simulated field keeping higher-order terms, and an ideal parabola given by $B_{ap} = C^{(0)} + x^2$, where $C^{(0)}$ is found using Eq.~(\ref{eq:Cn_even}) for the wire spacings.  The solid red curve shows the difference in the optimal wire configuration, and the dashed blue curve shows the difference in the low-power configuration.  The low-power configuration produces a field that is about an order of magnitude ``less harmonic'' than the wires in the optimal configuration.  However, for sufficiently small atomic clouds, $\sigma_{\parallel}/H < 0.05$, where $\sigma_{\parallel}$ is the axial size of the atomic cloud, both configurations produce potentials that are harmonic to one part in $10^{-9}$. 

To determine the amount of current that needs to be run in each wire of the chip, we need to determine the scaling of the current.  To make a harmonic trap, with trap frequency $\omega$, the scaling current should be
\begin{equation}
  \label{eq:scaling_current}
  I_R = \frac{\pi H^{3} m \omega^2}{\mu \mu_0},
\end{equation}
where, as before, $\mu$ is the magnetic moment of the trapped state.
Trapping $^{87}\mbox{Rb}$ in the $F=2$, $m_f = 2$ state in a trap with frequency $\omega = 2 \pi \times 10$~Hz that is $H = 1.6$~mm from the pinch wires, means that $I_R = 0.63$~A.  
Applying this scaling current, the currents in the high-power configuration are $I_{\mathcal{E},0} = 0.84$~A,  $I_{\mathcal{E},1} = 1.45$~A,  $I_{\mathcal{E},2H} = 7.79$~A, with a bottom field of $B_R C^{(0)} = 7.86$~G.   
For the low-power configuration, the currents are
$I_{\mathcal{E},0} = 0.20$~A,  $I_{\mathcal{E},1} = 0.18$~A,  $I_{\mathcal{E},2L} = 2.00$~A, and the bottom field is $B_R C^{(0)} = 2.59$~Gauss.   

\subsection{Double-well trap}

The same chip can be used to produce a double-well trap, where both the distance between the two traps and the difference between the potential at the bottom of each trap can be independently tuned.  This type of double-well trap can be used to study the merging of two cold or ultracold atomic clouds and the quantum dynamics of a Bose-Einstein condensate (BEC) in a double-well potential, or most interestingly, it may be useful as a coherent splitter for a BEC.  

\begin{figure}
\includegraphics[width=\columnwidth]{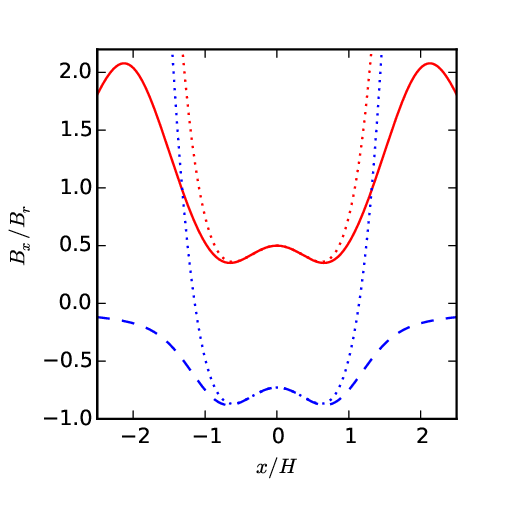}
\caption{A double-well potential created by tuning the terms $C^{(2)}$ and $C^{(4)}$. For both configurations $C^{(2)} = -0.75$ and $C^{(4)}= 1.0$.  In the optimal configuration (solid red line) the potential is deeper and has a large bottom field that may be offset with an external bias field.  In the low-power configuration (dashed blue line) the field is negative and would need to be offset with an external bias field to maintain the shape of the double well. 
\label{Fig:doubleWell}}
\end{figure}
Figure ~\ref{Fig:doubleWell} shows a double-well magnetic field produced by our chip.
The solid red curve in Fig.~\ref{Fig:doubleWell} shows the magnetic field produced by the pinch wires in the optimal configuration for a double-well trap with parameters $C^{(2)} = -0.75$, and $C^{(4)} = 1$.  The dashed blue curve is the field produced by the pinch wires in the low-power configuration.  The two dotted curves are the approximate values when no higher-order contributions to the field are included. Figure ~\ref{Fig:doubleWell} is an example of how two traps that have the same shape near the origin can have very different behavior far from the origin.  For the trap created using the wires in the optimal configuration, the bottom field is positive.  To reduce the size of this bottom field, a negative bias field must be applied. The field has a maximum before it tends towards zero.  On the other hand, for the low-power configuration, the field is always negative.  Since the absolute value of the field determines the potential, in order to create a double well, there must be a positive bias field applied to lift the field such that it is always positive.  The 
field has no other extrema and tends towards zero after the double-well structure.

\begin{figure}
\includegraphics[width=\columnwidth]{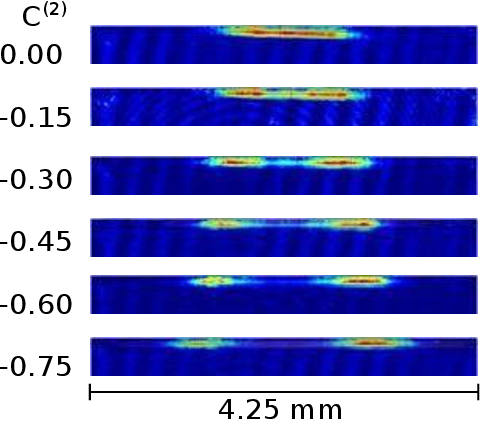}
\caption{{Experimental images of an atom cloud at appropriately 2~$\mu$K being transitioned from a pure $C^{(4)}= 1$ state to the double-well state shown in Fig.~\ref{Fig:doubleWell} as a function of the $C^{(2)}$ parameter.  There is an apparent tilt between the two wells that results in a number imbalance between the two wells.  It has not been determined if this is a physical tilt of the waveguide or an $\hat e_z$ gradient.} \label{Fig:splittingWell}}
\end{figure}

For the experimental realization of the double well, seen in Fig.~\ref{Fig:splittingWell}, the reference current, which determines an overall scaling of the total field, is set to $I_R = 1$~A. For a harmonic trap with $C^{(2)}=1$, this would produce a trap frequency of $\omega = 2 \pi \times 12.5$~Hz. However, the trap frequency at the minima of the two wells is reduced to approximately $\omega = 2 \pi \times 10$~Hz for this particular choice of the $C^{(2)}$ and $C^{(4)}$ parameters. The locations of the two wells are $x/H = \pm \sqrt{3/8}$ for an ideal potential of this form.   For the high-power configuration, the applied currents are $I_{\mathcal{E},0} = 0.47$~A, $I_{\mathcal{E},1} = -1.02$~A, and $I_{\mathcal{E},2H} = 2.27$~A.  For the low-power configuration, $I_{\mathcal{E},0} = 0.29$~A, $I_{\mathcal{E},1} = -1.39$~A, and $I_{\mathcal{E},2L} = 0.58$~A. Only the high-power configuration was investigated experimentally.

We show preliminary experimental results of a tunable atom chip well in Fig.~\ref{Fig:splittingWell}, where an approximately 2~$\mu$K atom cloud of $^{87}$Rb atoms in the $\left| F=2, m_F = 2\right>$ state is trapped on an atom chip similar to the chip shown in Fig.~\ref{fig:TuneableTrapWires}.  The pure fourth-order potential is modified by the addition of a negative $C^{(2)}$ contribution that splits the potential into two wells. Further results are being prepared for future publication.

\section{CONCLUSIONS}{\label{sec:Conclusions}}

We have demonstrated that tunability of an axial magnetic field in a cold-atom waveguide can be achieved with sets of paired wires on an atom chip. By symmetry, wires with (antiparallel) parallel currents contribute to only the (odd) even terms in the polynomial expansion of the field along the guide axis. When a wire pair is placed at a zero of a particular coefficient, it allows the lower-order terms (of the same parity) to be adjusted without contributing to the coefficient itself. Several wire pairs, appropriately placed, lead to arbitrary tunability of $N-1$ coefficients simply by controlling the relative currents through the sets of wire pairs. Experiments that employ 1D potentials
now have a tool with which precise potentials may be generated from a double-layer atom chip.   We have also shown the initial operation of a tunable atom chip by trapping a 2~$\mu$K cloud of $^{87}$Rb atoms in a pure fourth-order potential and in a double-well configuration that is composed of $C^{(2)}  = -0.75$  and $C^{(4)}= 1.0$.

\section*{Acknowledgment}

This work was funded under the Air Force Office of Scientific Research under program/task 10RV03COR.

\appendix

\section{Separability of the waveguide potential\label{sec:oneDField}}

A magnetic waveguide is a field configuration where the magnetic field vanishes
along an axis.  Near the zero, the field points perpendicularly to the guide and can be described by a
single parameter $G$, which is the magnetic-field gradient of the waveguide.  For example, the
magnetic field for a waveguide that points in the $x$ direction can be written as
\begin{equation}
  \label{eq:WaveGuide}
  B_{\mbox{radial}} = G ( \hat e_y y - \hat e_z z).
\end{equation}
The 1D potential will be created using a magnetic field that is produced by the current in several wires that run parallel to the $y$ axis (perpendicular to the waveguide axis).  This field will provide confinement in the axial direction and will be assumed to be of the form,
\begin{equation}
  \label{eq:AxialPotential}
  B_{\mbox{axial}} = B_x(x,z) \hat e_x + B_z(x,z) \hat e_z.
\end{equation}
The $z$ dependence in Eq.~(\ref{eq:AxialPotential}) causes two small shifts to the potential.  First, it causes a change in the gradient in the $z$ direction, which can be neglected when $G \gg \frac{\partial B_z}{\partial z}$.  Next, it causes a displacement in the $z$ direction, which can be neglected when $G^2 \sigma_x \gg \frac{\partial B_x^2}{\partial z} $ where $\sigma_x$ is the size of the cloud in the $x$ direction. When these inequalities are satisfied, Eq.~(\ref{eq:AxialPotential}) reduces to
  \begin{equation}
    \label{eq:axial2}
    B_{\mbox{axial}} = B_x^{T}(x) \hat e_x + B_z(x) \hat e_z.
  \end{equation}
The $x$ component of the magnetic field creates the potential along the waveguide and is the field that we wish to control.  The $z$ component of this field causes deformations to the waveguide.  The constant term $B_z^{(0)} = B_R D^{(0)}$ in Eq.~(\ref{eq:Bz}) causes a shift in the location of the guide by $B_z^{(0)}/G$
along the $y$ axis, 
which can be corrected using a uniform bias field in the $z$ direction. We assume that the appropriate zeroing bias is applied. The second term causes a rotation of the waveguide about $z$ in the 
$x-y$
plane.  The waveguide is rotated by the angle, $\theta \approx B_z^{(1)} / G$.  When using optical pulses to manipulate the state of the trapped atoms, this rotation angle becomes important.
Typically, this angle will be set to zero and neglected.  However, including nonzero rotations is straightforward.  The higher-order contributions to $B_z$ cause other distortions to the path of the waveguide, but below those effects will not be considered.  

With $B_{z}$ set to zero, the field along the waveguide axis can be separated into two parts: a non zero ``bottom field'' $B_x(0)$ which prevents spin-flip losses and is necessary for the potential to be separable and $B_x(x)$, the part of the axial field that depends on the $x$ coordinate,
\begin{equation}
  \label{eq:axial3}
  B_{\mbox{axial}} = B^{T}_{x}(x) = B_x(0) + B_x(x).
\end{equation}

The potential that the atoms experience is obtained from the radial and axial components, given by Eqs.~(\ref{eq:WaveGuide}) and (\ref{eq:axial3}), respectively, as follows:
\begin{equation}
  \label{eq:Potential}
  V = \mu \sqrt{ \left[ B_x(0) + B_x(x)\right]^2 + G^2 r_\perp^2},
\end{equation}
where $\mu$ is the magnetic moment of the trapped state and
$r_\perp = \sqrt{y^2 +  z^2}$ is the radial coordinate.

Assuming that $B_x(0) \gg B_x(x)$ and expanding Eq.~(\ref{eq:Potential}) yields
\begin{equation}
  \label{eq:PotentialExpansion}
  \begin{split}
  V =& \mu \left(
    |B_x(0) + B_x(x)| + \frac{1}{2} \frac{G^{2}}{|B_x(0)|} r_\perp^2 \right.\\ &- \left. \frac{1}{2} \frac{G^{2}}{|B_x(0)|B_x(0)} B_x(x) r_\perp^2
 \right).
 \end{split}
\end{equation}
The last term in  Eq.~(\ref{eq:PotentialExpansion}) is clearly not separable; that is, it cannot
be written in the form $V = V_{\mbox{axial}}(x) + V_{\mbox{radial}}(r_\perp)$.  However, the potential may be regarded as separable in the limit where $B_x(0)^2 \gg G^2 \sigma_{\perp}^2$, where $\sigma_{\perp}$ is the size of the atomic cloud in the radial direction.

From Eq. (\ref{eq:PotentialExpansion}) it is clear 
that the potential along the waveguide can be written in the form shown in Eq.~(\ref{eq:Ve}).

\section{Derivation of wire coefficients\label{sec:GenExp}}

Take the surface of the atom chip to be at $z=0$, with an infinitely long wire parallel to the $y$ axis along the line $x^*=W$. When a current of $I_{R}$ is passed through the wire, the $x$ component of a magnetic field a distance $z=H$ above the atom chip, at field point $X$, is given by, 
\begin{equation}
  \label{eq:Bprop}
  B_x(x) = B_R \frac{1}{1 + (x-w)^2},
\end{equation}
where $x = X/H$, and $w = W/H$, and $B_R=\mu_0 I_{R} / 2 \pi H$.
Equation (\ref{eq:Bprop}) can be expanded as the series
\begin{equation}
  \label{eq:Bprop2}
  B_x(x)/B_R = \sum_{n=0}^{\infty} \sum_{k=0}^{2n} (-1)^{(n+k)} \binom{2n}{k} w^{2n-k}x^{k},
\end{equation}
To determine the coefficients for each power of $x$, the order of the summation in Eq.~(\ref{eq:Bprop2}) needs to be interchanged.  To do this first, the even and odd terms are separated so that the upper limit of the inner summation can be divided in half, i.e. $\sum_{k=0}^{2n} A_{n,k} = \sum_{k=0}^n (A_{n,2k} + A_{n+1, 2k+1})$.  Then, the order summation can be flipped $\sum_{n=0}^{\infty} \sum_{k=0}^n B_{n,k} = \sum_{k=0}^{\infty} \sum_{n=k}^{\infty} B_{n,k} = \sum_{k=0}^{\infty} \sum_{n=0}^{\infty} B_{k+n, k}$.   Finally, after interchanging the order of the summation, Eq.~(\ref{eq:Bprop2}) becomes 
\begin{equation}
  \label{eq:Bprop3}
  B_x(x)/B_R  = \sum_{k=0}^{\infty} (\alpha^{2k} x^{2k} + \alpha^{2k+1} x^{2k+1}),
\end{equation}
where
\begin{equation}
  \label{eq:an}
  \alpha^{(n)} = \sum_{q=0}^{\infty} (-1)^{(n+q)/2} \binom{n+q}{n} w^{q} \phi_{n+q},
\end{equation}
where $\phi_{a} = [1+(-1)^a]/2$ is $0$ when $a$ is odd and $1$ when $a$ is even.  

Using the identity
\begin{equation}
  \label{eq:binomRelation}
  \binom{n+q}{n} = \sum_{r=0}^{r_{max}} \binom{n + (q-r)/2}{n} \binom{n+1}{r} \phi_{q+r},
\end{equation}
where $r_{max} = \min(q, n+1)$, it is assumed that both $q$ and $n$ are positive integers. 
Substituting (\ref{eq:binomRelation}) into (\ref{eq:cn}) and reversing the order of the summation yields
\begin{equation}
  \label{eq:cn2}
  \begin{split}
  \alpha^{(n)} = \sum_{r=0}^{n+1} \sum_{q=r}^{\infty} & (-1)^{(n+q)/2} \binom{n + (q-r)/2}{n} \binom{n+1}{r} \\ & \times w^{q} \phi_{q+r} \phi_{n+q}.
  \end{split}
\end{equation}
Eq.~(\ref{eq:cn2}) can be written as the product of two sums, by introducing the new index,
$\kappa = (q-r)/2$ resulting in 
\begin{equation}
  \label{eq:cn3}
  \begin{split}
  \alpha^{(n)} = &\left[ \sum_r (-1)^{(n+r)/2} \binom{n+1}{r} w^{r} \phi_{n+r} \right] \\ \times & \left[ \sum_{\kappa}
(-1)^{\kappa} \binom{n+\kappa}{n} w^{2 \kappa}
 \right].
 \end{split}
\end{equation}
Recognizing that the second term in Eq.~(\ref{eq:cn3}) can be written as $(1+w^{2})^{-n-1}$, we can write the coefficients as
\begin{equation}
  \label{eq:cn4}
  \alpha^{(n)}(w) = \frac{1}{(1+w^2)^{n+1}} \sum_{r=0}^{n} (-1)^{(n+r)/2} \binom{n+1}{r} w^{r} \phi_{n+r}.
\end{equation}

Each of the wires contributes to all of the coefficients.  Contributions to the magnetic field with definite parity can be created using pairs of wires.   A pair of wires will be located at $\pm w$.  If the current is running in the same (opposite) direction, only even (odd) terms will contribute to the potential.  For a pair of wires, the coefficients will be larger by a factor of $2$, i.e., $c^{(n)} = 2 \alpha^{(n)}$.  

\bibliography{TuneableTrap}

\end{document}